\documentclass[graybox]{svmult}

\usepackage{mathptmx}       % selects Times Roman as basic font
\usepackage{helvet}         % selects Helvetica as sans-serif font
\usepackage{courier}        % selects Courier as typewriter font
\usepackage{type1cm}        % activate if the above 3 fonts are
\usepackage{makeidx}         % allows index generation
\usepackage{graphicx}        % standard LaTeX graphics tool
\usepackage{multicol}        % used for the two-column index
\usepackage[bottom]{footmisc}% places footnotes at page bottom

\makeindex             % used for the subject index
\begin{document}

\title*{From fields to a super-cluster: the role of the environment at z=0.84 with HiZELS}
\author{David Sobral, Philip Best, Ian Smail, Jim Geach \& HiZELS team}
\institute{David Sobral \at Institute for Astronomy, ROE, Blackford Hill, Edinburgh, \email{drss@roe.ac.uk}}
\maketitle

% Too much empty space in the original style file!
\vskip-1.3truein

\abstract{At $z=0$, clusters are primarily populated by red, elliptical and massive galaxies, while blue, spiral and lower-mass galaxies are common in low-density environments. Understanding how and when these differences were established is of absolute importance for our understanding of galaxy formation and evolution, but results at high-$z$ remain contradictory. By taking advantage of the widest and deepest H$\alpha$ narrow-band survey at $z=0.84$ over the COSMOS and UKIDSS UDS fields, probing a wide range of densities (from poor fields to rich groups and clusters, including a confirmed super-cluster with a striking filamentary structure), we show that the fraction of star-forming galaxies falls continuously from $\sim40$\% in fields to approaching 0\% in rich groups/clusters. We also find that the median SFR increases with environmental density, at least up to group densities -- but only for low and medium mass galaxies, and thus such enhancement is mass-dependent at $z\sim1$. The environment also plays a role in setting the faint-end slope ($\alpha$) of the H$\alpha$ luminosity function. Our findings provide a sharper view on galaxy formation and evolution and reconcile previously contradictory results at $z\sim1$: stellar mass is the primary predictor of star formation activity, but the environment also plays a major role.}

\vspace*{-0.38cm}

\section{Introduction}
\label{sec:1}

\vspace*{-0.18cm}
Star formation activity is strongly dependent on environment: clusters of galaxies are primarily populated by passively-evolving galaxies, while star-forming galaxies are common in low-density environments \cite{ad80}. It is well-established \cite{BEST04} that the typical star formation rates of galaxies -- and the star-forming fraction -- decrease with local galaxy density (often projected local density, $\Sigma$) both in the local Universe and at moderate redshift \cite{KODAMA04}. Active star-forming galaxies in the local Universe are also found to have lower masses than passive galaxies and, indeed, the most massive galaxies are mostly non-star-forming, an observational result often known as mass-downsizing \cite{COWIE}. While massive galaxies are predominantly found in high density environments, it has been shown that the mass-downsizing trend is not simply a consequence of the environmental dependence, nor vice-versa \cite{PENG}.

\begin{figure}[b]
\sidecaption
% For example, with the graphicx style use
\includegraphics[scale=.73]{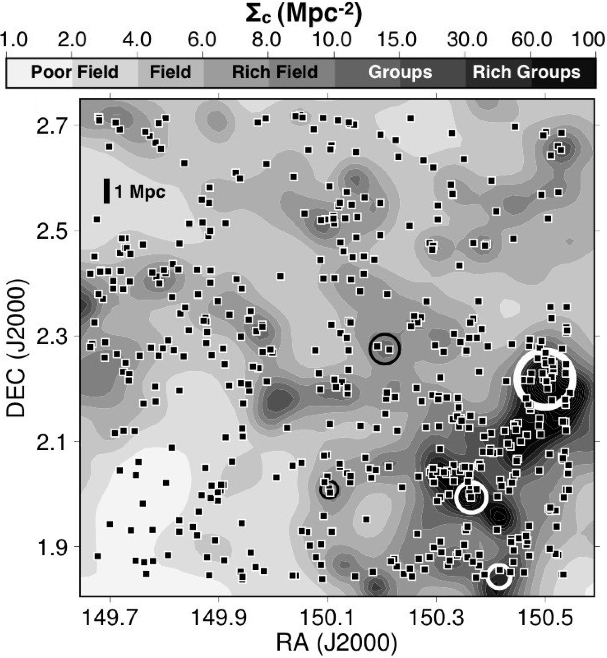}
\caption{The on-sky distribution of star-forming H$\alpha$ emitters at $z=0.84$ in the COSMOS field (filled squares), compared with the local projected density field (grayscale; using a 10th nearest neighbor analysis). Circles mark the positions of extended X-ray emission from confirmed groups and clusters within the narrow-band redshift range, scaled to reflect the measured X-ray luminosity (in a log scale). Lower X-ray luminosity groups are identified with black circles, while richer X-ray groups are plotted in white for good contrast. Note the very rich structure in the COSMOS field (including a rich cluster), providing a unique opportunity to probe the densest environments.}
\label{FIG1}       % Give a unique label!
\end{figure}

When did the environment and mass dependences of star-forming galaxies start to be observable, and how did they affect the evolution of galaxies and clusters? By $z\sim1$, some authors \cite{ELBAZ,COOPER} have claimed to have found a flattening or even a definitive reverse of the relation between star formation activity and local galaxy density (SFR-$\Sigma$ relation). However, other studies \cite{FINN, POGGIANTI,PATEL} argue that even at $z\sim1$ both star formation rate and the star-forming fraction decline with increasing local density. Part of the discrepancies may be due to mass dependences already in place at high redshift. In order to identify and distinguish the separate roles of mass and environment on star formation at high redshift, one really requires clean, robust and large samples of star-forming galaxies residing in a wide range of environments and with a wide range of masses, together with samples of the underlying population found at the same redshift. Narrow-band H$\alpha$ surveys are one of the most effective ways to gather representative samples of star-forming galaxies at different epochs, and the scientific potential of these is now being widely explored, following the development of wide-field cameras in the near-infrared. In particular, HiZELS, the Hi-Redshift($z$) Emission Line Survey \cite{GEACH2008,SOBRAL09A}, is playing a world-leading role by obtaining very large samples of H$\alpha$ emitters (and other emission lines; see \cite{SOBRAL09B}), at $z=0.84$, $z=1.47$ and $z=2.23$ (see \cite{BEST2010} for an overview) over various square degree fields with a wealth of high-quality multi-wavelength data.

%Remember the page limits for Symposium 2 of JENAM 2010 are as follows:
%\begin{itemize}
%\%item Invited Talks: 8 pages
%\item Contributed Talks: 5 pages
%\item Posters: 2 pages
%\end{itemize}

%The deadline is November 20th, 2010. Your contributions should be sent to {\tt jenam2010gf@mssl.ucl.ac.uk} (\LaTeX file and figures), and you should also send us a signed copy of the agreement form, found in our webpages  \url{www.mssl.ucl.ac.uk/$\sim$ipf/GF.html} 

\vspace*{-0.38cm}

\section{SAMPLES AND PROPERTIES}
\label{sec:2}
% Always give a unique label
% and use \ref{<label>} for cross-references
% and \cite{<label>} for bibliographic references
% use \sectionmark{}
% to alter or adjust the section heading in the running head
\vspace*{-0.18cm}

%subsection{Star-forming population at $z=0.84$}

This study uses the large sample of H$\alpha$ emitters at $z=0.84$ from HiZELS presented in \cite{SOBRAL09A}, modified as detailed in \cite{SOBRAL10B}. Briefly, the sample was derived from a narrow-band $J$ filter ($\Delta\lambda=0.015\umu$m) survey using WFCAM/UKIRT and reaching a star formation rate (SFR) limit in H$\alpha$ of 3 M$_{\odot}$\,yr$^{-1}$ over $1.3\deg^2$ in the UKIDSS UDS and the COSMOS fields. Photometric redshifts were used to select a sample which is $>95$\% reliable and complete (based on zCOSMOS - see \cite{SOBRAL09A}); this has been further modified to i) take into account highly improved photometric redshifts, ii) reject potential AGN identified in \cite{GARN2010} and iii) include EW$<50$\,\AA \ H$\alpha$ emitters (c.f. \cite{SOBRAL10B}). The final sample contains a total of 770 star-forming H$\alpha$ emitters.

The improved high quality photometric redshifts at $z\sim0.8$ available in COSMOS and UDS are used to emulate the narrow-band filter selection and to select the underlying population. Spectroscopic redshifts provide completeness and contamination estimates of various photometric-redshift selected samples, allowing for appropriate corrections to be made. The underlying sample contains 6344 sources.

Stellar masses are determined using a detailed SED fitting (with a wide range of parameters and fixing $z=0.84$), and environmental densities are based on a 10th nearest neighbor analysis, corrected for completeness and contamination by using spectroscopic redshifts (c.f. \cite{SOBRAL10B}, including detailed studies showing how the results are robust against errors and systematics). Local environmental densities are classified using the real-space correlation length, using the analysis presented in \cite{SOBRAL10A} into $fields$ ($\Sigma_{\rm c}<10$\,Mpc$^{-2}$), $groups$ ($10<\Sigma_{\rm c}<30$\,Mpc$^{-2}$) and \it rich groups/clusters \rm ($\Sigma_{\rm c}>30$\,Mpc$^{-2}$). A very wide range of environments is probed, confirmed by the detection of extended X-ray emission in the highest density regions (see Fig 1).

\begin{figure}[b]
\sidecaption
% For example, with the graphicx style use
\includegraphics[scale=.74]{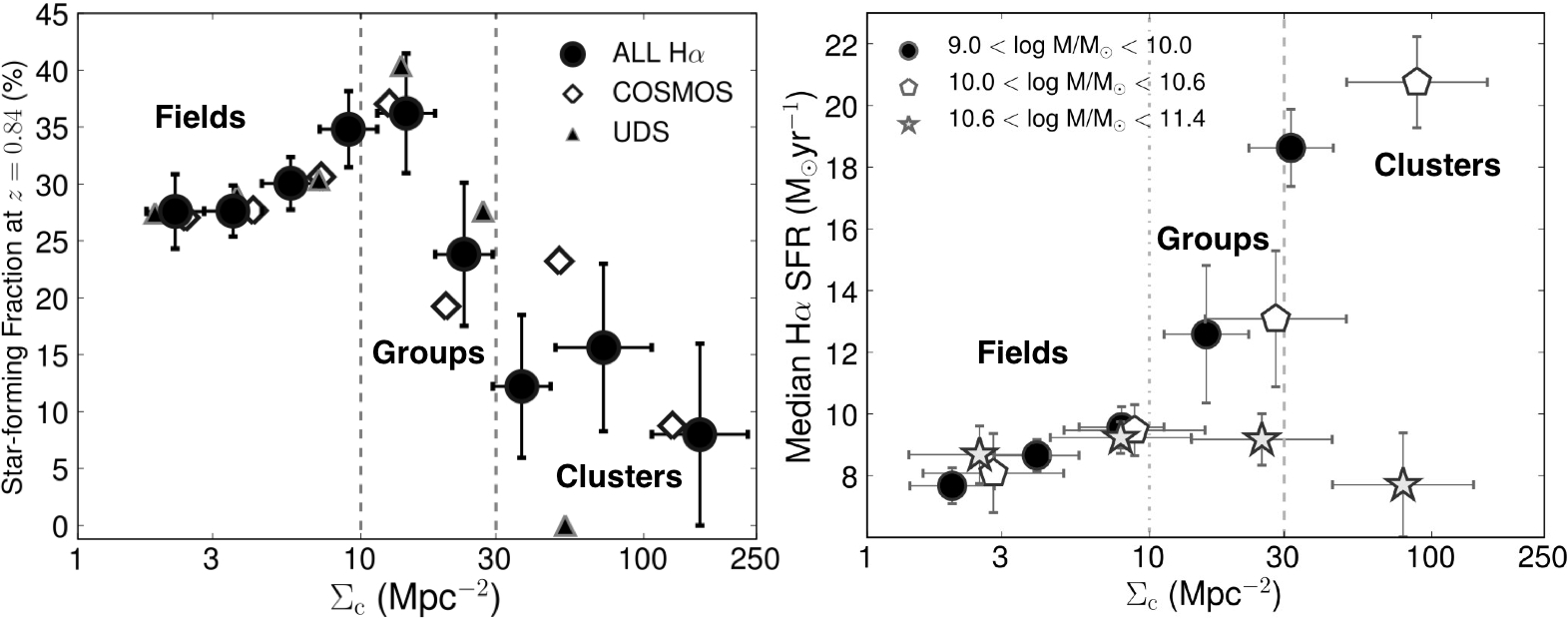}
\caption{The fraction of the H$\alpha$ star-forming galaxies (left panel) and the median H$\alpha$ star formation rate of H$\alpha$ emitters in different stellar mass bins (right panel), as a function of local projected galaxy surface density (SFRs\,$>3$\,M$_{\odot}$\,yr$^{-1}$). At the lowest densities, the star-forming fraction is relatively constant and only increases slightly with $\Sigma_{\rm c}$. However, for higher densities, $\Sigma_{\rm c}>10$\,Mpc$^{-2}$, there is a steep decline of the star-forming fraction down to the highest (rich groups/cluster) densities probed. The results also show that the typical (median) SFR of H$\alpha$ emitters increases continuously from the lowest densities to group densities, but this is only found for low/medium masses: massive star-forming galaxies have SFRs which are mostly unaffected by their environment.}
\label{FIG2}       % Give a unique label!
\end{figure}

\vspace*{-0.38cm}

\section{ENVIRONMENTAL DEPENDENCES AT $\bf Z\sim1$}
\label{sec:3}

\vspace*{-0.18cm}

We find that the fraction of galaxies forming stars (above the HiZELS limit) is relatively flat with increasing local density within the field regime, but it falls sharply with density once group densities are reached (Fig.2), resulting in a fall from the field to rich groups/clusters, as seen in the nearby Universe and consistent with \cite{FINN, PATEL}. The median star formation rates of star-forming galaxies increases with environmental density for both field and group environments (Fig.2), in good agreement with \cite{ELBAZ,COOPER}, but the trend is stopped for the highest densities, where it is appears to fall (c.f. \cite{SOBRAL10B}), also agreeing with studies probing such very high densities \cite{POGGIANTI}. Furthermore, we find that the environment changes the shape of the H$\alpha$ luminosity function: the faint-end slope ($\alpha$) is found to vary with environmental density, being very steep ($\alpha\sim-2$) for poor fields, and very shallow for the highest density regions (groups and clusters, $\alpha\sim-1$), as shown in Fig. 3.

\begin{figure}[h]
\sidecaption
% For example, with the graphicx style use
\includegraphics[scale=.385]{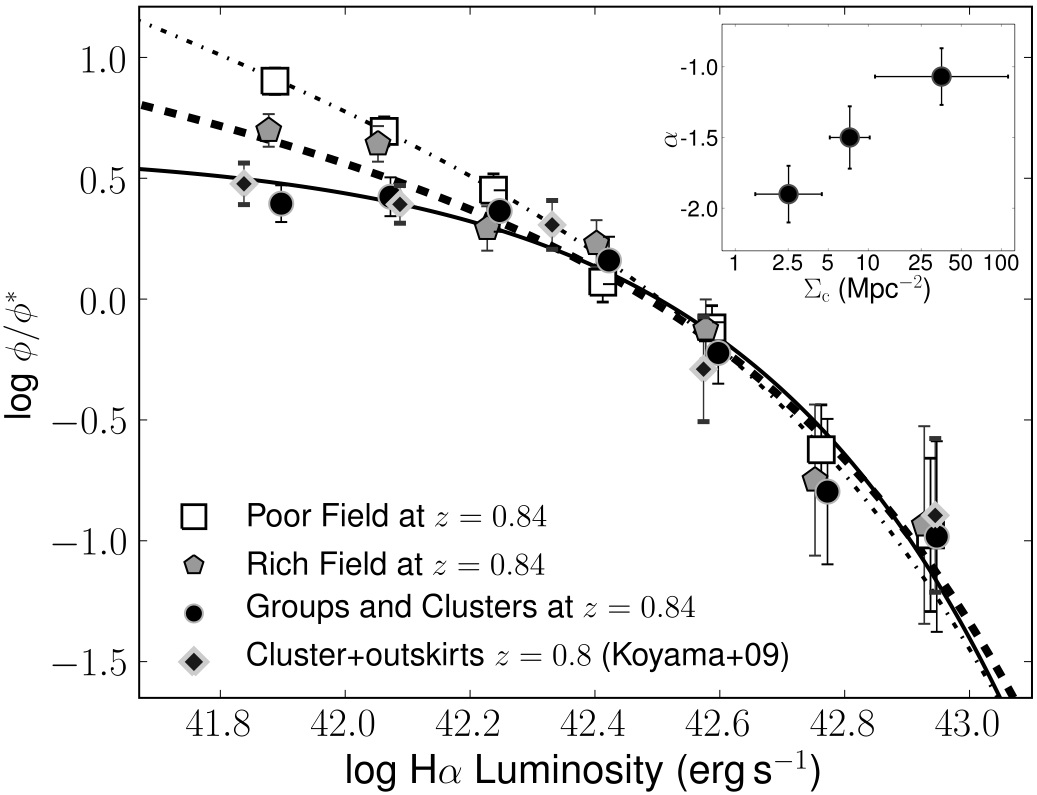}
\caption{The normalised H$\alpha$ luminosity function shows a significant difference in the faint-end slope, $\alpha$, as a function of local surface density. As shown in the inset, $\alpha$ is very steep ($\alpha\approx-1.9$) for poor field, shallower for medium densities (rich field), and very shallow for groups and clusters, with $\alpha\approx-1.1$. Recent results from \cite{KOYAMA} fully agree with the H$\alpha$ luminosity function derived for similar local projected densities.}
\label{FIG3}       % Give a unique label!
\end{figure}

\vspace*{-0.38cm}

\section{MASS-ENVIRONMENT VIEW AT $\bf Z\sim1$}
\label{sec:4}

\vspace*{-0.18cm}

We find that mass-downsizing is fully in place at $z\sim1$, with the fraction of star-forming galaxies declining steeply with stellar mass (c.f. \cite{SOBRAL10B}). Since stellar mass and environment are correlated, to which extent could the environmental trends be driven by stronger, more fundamental mass trends? In order to address this question, we have taken advantage of the large samples to investigate dependences on both mass and environment, simultaneously. Interestingly, when potentially merger-driven star-formation (which dominates at high densities) is neglected, we find that the fraction of star-forming galaxies declines independently with both environment and stellar mass (fixing the other), clearly showing no qualitative evolution from the local Universe (Fig. 4). However, we also find that the $\sim4000$\,\AA\, break colour of star-forming galaxies depends almost uniquely, and very strongly, on stellar mass (Fig.4), and not on environment. Interestingly, we also find that the positive SFR-$\Sigma$ correlation is driven by low and median stellar mass galaxies: the median star formation rates of the most massive galaxies are mostly unaffected by the environment, and $s$SFRs of such massive galaxies actually decline with $\Sigma$; this fully explains apparently contradictory results in the literature which used different selections and probed different environments.

Overall, we find that stellar mass is the primary predictor of star formation activity at $z\sim1$, but the environment, while initially enhancing the star formation activity of (lower-mass) star-forming galaxies, is ultimately responsible for suppressing star-formation activity in all galaxies above surface densities of groups and clusters.

%References should be \textit{cited} in the text by 
%number.\footnote{Make sure that all references
%from the list are cited in the text.} The
%reference list should ideally be \textit{sorted} in alphabetical order
%-- even if reference numbers are used for the their citation in the
%text. If there are several works by the same author, the following
%order should be used:
%\begin{enumerate}
%\item all works by the author alone, ordered chronologically by year of publication
%\item all works by the author with a coauthor, ordered alphabetically by coauthor
%\item all works by the author with several coauthors, ordered chronologically by year of publication.
%\end{enumerate}

%Regarding the \textit{styling} of references\footnote{Always use the
 % standard abbreviation of a journal's name according to the ISSN
  %\textit{List of Title Word Abbreviations}, see
 % \url{http://www.issn.org/en/node/344}}, we will follow the standard
%of Monthly Notices of the RAS, e.g.~\cite{ad80,if09,ap10}.

\begin{figure}[h]
\sidecaption
% For example, with the graphicx style use
\includegraphics[scale=.565]{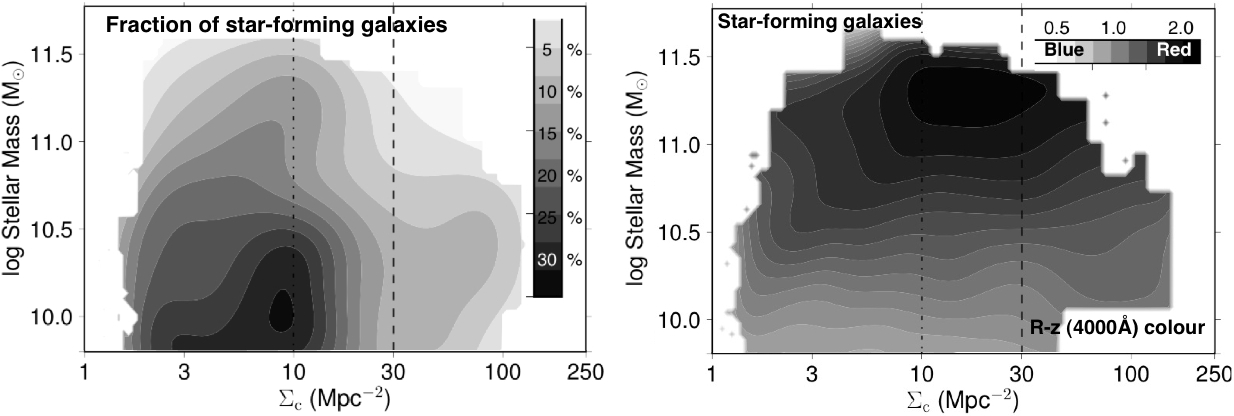}
\caption{$Left$: The fraction of star-forming galaxies (mergers excluded) as a function of both mass and environment, revealing that both are important at $z\sim1$, just like in the local Universe. $Right$: The distribution of the median R-z colour (roughly probing the 4000\,\AA \, break colour at $z=0.84$) within the mass-density 2D space for the H$\alpha$ star-forming galaxies; this reveals that stellar mass is the main colour predictor, as the environment only correlates weakly with colour.}
\label{FIG4}       % Give a unique label!
\end{figure}

\vspace*{-0.88cm}

%%% Bibliography

%\input{referenc}
\end{document}